\documentclass[11pt]{article}
\usepackage{amssymb,amsmath,amsfonts}
\usepackage{graphicx}
\usepackage{graphics}
\usepackage{eepic,epsfig}

\begin{document}

\title{Generalized Abel-Plana Formula as a Renormalization Tool \\
in Quantum Field Theory}
\author{A. A. Saharian\thanks{%
E-mail: saharian@ysu.am } \\
\textit{Institute of Physics, Yerevan State University, }\\
\textit{1 Alex Manogian Street, 0025 Yerevan, Armenia }}
\maketitle

\begin{abstract}
In quantum field theory the vacuum expectation values of physical
observables, bilinear in the field operator, diverge. Among the most
important points in the investigations of those expectation values is the
regularization of divergent expressions, separation of divergences and the
renormalization. In problems with boundaries the expectation values are
expressed in the form of the difference of the divergent series and the
corresponding integral. In problems with planar boundaries a finite integral
representation for that difference is provided by the Abel-Plana summation
formula. In the present contribution we consider the generalization of the
Abel-Plana formula that allows to obtain similar representations for more
general classes of series where the summation goes over the zeros of a given
function. Applications are discussed in quantum field theoretical problems
with nontrivial spatial topology and curved boundaries.
\end{abstract}

\bigskip

\\


\bigskip

\section{Introduction}

In quantum field theory the operators of a number of important physical
characteristics are expressed in terms of bilinear products of the field
operator or its spacetime derivatives evaluated at the same spacetime point.
Examples are the field squared, current density for charged fields and the
energy-momentum tensor. The corresponding expectation values for a given
quantum state are divergent and for the extraction of finite physical
results a regularization procedure with the subsequent renormalization is
required. For quantum fields in curved spacetime the structure of
divergences in local physical observables at a given point is determined by
the local geometric characteristics of the spacetime in the form of various
combinations of the Riemann tensor and its derivatives (see, for example, 
\cite{Birr82}). In particular, the local divergences are the same in two
problems considering the physical system in two spacetimes with the same
local but different global geometries. The difference can be induced, for
example, by introducing additional boundaries or by compactification of
spatial dimensions. The finite shift in the expectation values of local
physical characteristics, induced by the change of global geometry, is
expressed in terms of the difference of two divergent expressions
corresponding to initial and modified geometries. One of the ways to find
the finite difference is to regularize both the terms in the subtraction
procedure, evaluate the difference of the regularized expectation values and
then remove the regularization in the final expression. Different
regularization procedures have been considered in the literature, including
the introduction of the cutoff function in the sum over the field modes,
dimensional regularization and the zeta function technique. Of course, the
final result should not depend on the specific regularization scheme. Among
the most efficient ways to extract the finite contributions is the
application of the Abel-Plana summation formula (APF) (see, e.g., \cite%
{Henr74,Hard91,Saha08}). In particular, it has been extensively used in the
investigations of the Casimir effect for planar boundaries \cite%
{Most97,Bord09}. In the present contribution we discuss generalizations of
the APF obtained from the generalized Abel-Plana formula (GAPF) considered
in \cite{Saha87} (see also \cite{Saha08} for a review).

\section{Models with compact dimensions}

\label{sec:Topology}

In the discussion below, the physical applications of the GAPF will be
demonstrated for a quantum scalar field $\varphi (x)$ with mass $m$ and
curvature coupling parameter $\xi $. The corresponding dynamics is governed
by the Klein-Gordon equation (the system of units $\hbar =c=1$ is used) 
\begin{equation}
\left( g^{ik}\nabla _{i}\nabla _{k}+m^{2}+\xi R\right) \varphi (x)=0,
\label{Eqg}
\end{equation}%
where $g^{ik}$ is the metric tensor for the background spacetime, $\nabla
_{i}$ is the covariant derivative operator, and $R$ is the Ricci scalar. In
this section we consider a flat spacetime with the line element $%
ds^{2}=g_{ik}dx^{i}dx^{k}=dt^{2}-d\mathbf{x}^{2}$ covered by the Cartesian
spatial coordinates $\mathbf{x}=(x^{1},x^{2},\ldots ,x^{D})$. For this
special case, in (\ref{Eqg}) one has $\nabla _{i}=\partial _{i}=\partial
/\partial x^{i}$ and $R=0$. It will be assumed that the space has the
topology $R^{D-1}\times S^{1}$ with compact dimension $x^{D}$ of the length $%
2\pi L$, $0\leq x^{D}\leq 2\pi L$. For the remaining coordinates we take $%
-\infty <x^{l}<\infty $, $l=1,2,\ldots ,D-1$. In models with nontrivial
topology, in addition to (\ref{Eqg}), the periodicity condition along
compact dimensions needs to be specified. Here, the quasiperiodicity
condition 
\begin{equation}
\varphi (t,\mathbf{x}_{\perp },x^{D}+L)=e^{2i\pi \alpha }\varphi (t,\mathbf{x%
}_{\perp },x^{D}),  \label{Pcond}
\end{equation}%
will be imposed with a constant phase $\alpha $, where $\mathbf{x}_{\perp
}=(x^{1},x^{2},\ldots ,x^{D-1})$ stands for the set of the coordinates in
the uncompact subspace. Without loss of generality we can assume that $%
\left\vert \alpha \right\vert \leq 1/2$. The special cases $\alpha =0$ and $%
|\alpha |=1/2$ correspond the untwisted and twisted fields, most frequently
considered in the literature.

The condition (\ref{Pcond}) on the field operator modifies the spectrum of
vacuum fluctuations compared to the fluctuations in the problem where the
direction $x^{D}$ has trivial topology $R^{1}$ with $-\infty <x^{D}<+\infty $%
. As a consequence, the vacuum expectation values (VEVs) of physical
observables are shifted by an amount depending on the compactification
length and the phase. This is the manifestation of the topological Casimir
effect widely considered in the literature for different local geometries
and spatial topologies \cite{Most97,Bord09,Casi11}. The VEVs of the
observables bilinear in the field operator are obtained from the two-point
functions. Here we consider the Hadamard function (HF) defined as the VEV $%
G(x,x^{\prime })=\left\langle 0\right\vert \varphi (x)\varphi ^{\dagger
}(x^{\prime })+\varphi ^{\dagger }(x^{\prime })\varphi (x)\left\vert
0\right\rangle $, where $\left\vert 0\right\rangle $ is the vacuum state.
Expanding the field operator over a complete set of mode functions $\varphi
_{\sigma }^{(\pm )}(x)$, obeying the field equation (\ref{Eqg}) and the
condition (\ref{Pcond}), the HF is presented as the mode-sum $G(x,x^{\prime
})=\sum_{\sigma }\sum_{s=\pm }\varphi _{\sigma }^{(s)}(x)\varphi _{\sigma
}^{(s)\ast }(x^{\prime })$. The collective set of quantum numbers $\sigma $
specifies the solutions. In accordance with the problem symmetry it is
natural to take the modes with definite values of the momentum $\mathbf{p}%
=(p^{1},p^{2},\ldots ,p^{D})$. The normalized positive (upper sign) and
negative (lower sign) energy modes have the plane-wave form $\varphi
_{\sigma }^{(\pm )}(x)=e^{i\mathbf{p}\cdot \mathbf{x}\mp iE_{\mathbf{p}}t}/%
\sqrt{2(2\pi )^{D}LE_{\mathbf{p}}}$, with the energy $E_{\mathbf{p}}=\sqrt{%
\mathbf{p}^{2}+m^{2}}$, where $\mathbf{x}=(t,\mathbf{x}_{\perp },x^{D})$ and 
$\mathbf{p}\cdot \mathbf{r}=\sum_{l=1}^{D}p^{l}x^{l}$. The eigenvalues of
the component $p^{D}$ are determined from (\ref{Pcond}) and are given by $%
p^{D}=p_{n}\equiv (n+\alpha )/L$, $n=0,\pm 1,\pm 2,\ldots $. For the
remaining components we have $-\infty <p^{l}<+\infty $, $l=1,2,\ldots ,D-1$.
The set of quantum numbers is specified as $\sigma =(\mathbf{p}_{\perp },n)$
with $\mathbf{p}_{\perp }=(p^{1},p^{2},\ldots ,p^{D-1})$.

Plugging the mode functions in the sum over the modes we get%
\begin{equation}
G(x,x^{\prime })=\frac{1}{L}\int \frac{d\mathbf{p}_{\perp }}{(2\pi )^{D}}e^{i%
\mathbf{p}_{\perp }\cdot \Delta \mathbf{x}_{\perp }}\sum_{n=-\infty
}^{+\infty }\frac{e^{ip_{n}\Delta x^{D}}}{E_{\mathbf{p}}}\cos (E_{\mathbf{p}%
}\Delta t),  \label{G2}
\end{equation}%
with $\Delta \mathbf{x}_{\perp }=\mathbf{x}_{\perp }-\mathbf{x}_{\perp
}^{\prime }$, $\Delta x^{D}=x^{D}-x^{\prime D}$, and $\Delta t=t-t^{\prime }$%
. The HF $G_{0}(x,x^{\prime })$ in the problem with trivial spatial topology 
$R^{D}$ is given by%
\begin{equation}
G_{0}(x,x^{\prime })=\int \frac{d\mathbf{p}}{(2\pi )^{D}}\frac{e^{i\mathbf{p}%
\cdot \Delta \mathbf{x}}}{E_{\mathbf{p}}}\cos (E_{\mathbf{p}}\Delta t).
\label{G0}
\end{equation}%
The VEVs of the observables bilinear in the field operator are obtained from
the Hadamard function or its derivatives in the coincidence limit $x^{\prime
}\rightarrow x$ of the arguments. That limit is divergent. An example is the
VEV of the field squared, $\left\langle 0\right\vert \varphi (x)\varphi
^{\dagger }(x)\left\vert 0\right\rangle =\left\langle \varphi \varphi
^{\dagger }\right\rangle $, given as $\left\langle \varphi \varphi ^{\dagger
}\right\rangle =\lim_{x^{\prime }\rightarrow x}G(x,x^{\prime })/2$. The
effects induced by the compactification are encoded in the difference $%
\left\langle \varphi \varphi ^{\dagger }\right\rangle -\left\langle \varphi
\varphi ^{\dagger }\right\rangle _{0}$, where $\left\langle \varphi \varphi
^{\dagger }\right\rangle _{0}$ is the VEV for the geometry with trivial
topology and is obtained from (\ref{G0}) in the coincidence limit. The local
geometry for the topologies $R^{D-1}\times S^{1}$ and $R^{D}$ is the same
and the difference $\left\langle \varphi \varphi ^{\dagger }\right\rangle
-\left\langle \varphi \varphi ^{\dagger }\right\rangle _{0}$ is finite.

An integral representation of the series in (\ref{G2}), convenient for the
separation of the effects of nontrivial topology, is obtained by using the
GAPF (see formula (2.11) in \cite{Saha08}). The latter contains two
functions $g(z)$ and $f(z)$, meromorphic in the right half-plane of the
complex variable $z$. We fix the function $g(z)$ by the relation 
\begin{equation}
g(z)=-if(z)\cot \pi \left( z+\beta \right) ,\;0<\beta <1,  \label{gz}
\end{equation}%
with the function $f(z)$ analytic in the right half-plane. The function $g(z)
$ has simple poles in the right half-plane located at $z=z_{g,n}=n-\beta $
with $n=1,2,\ldots $. Taking the limit $a\rightarrow 0$ in the GAPF we get%
\begin{equation}
\lim_{b\rightarrow \infty }\left[ \sum_{n=1}^{n_{b}}f(n-\beta )-\int_{0}^{b}{%
dx\,f(x)}\right] =i\int_{0}^{\infty }{dx\,\sum_{s=\pm 1}\,\frac{sf(e^{si\pi
/2}x)}{e^{2\pi \left( x-si\beta \right) }-1},}  \label{SF1}
\end{equation}%
where $n_{b}$ is defined by the condition $n_{b}\leq b+\beta <n_{b}+1$. The
corresponding formula for $-1<\beta <0$ is obtained by using the relation $%
\sum_{n=0}^{n_{b}}f(n-\beta )=\sum_{n=1}^{n_{b}+1}f(n-\beta ^{\prime })$,
with $\beta ^{\prime }=1+\beta $, and applying (\ref{SF1}). The only
difference from (\ref{SF1}) is that now the summation in the left-hand side
goes in the range $0\leq n\leq n_{b}$. If both the series and integral in (%
\ref{SF1}) are convergent, the combined formula is written as%
\begin{equation}
\sum_{n=n_{0}}^{\infty }f(n-\beta )=\int_{0}^{\infty }{dx\,f(x)+}%
i\int_{0}^{\infty }{dx\,}\sum_{s=\pm 1}{\frac{sf(e^{si\pi /2}x)}{e^{2\pi
\left( x-si\beta \right) }-1},}  \label{SF2}
\end{equation}%
where $n_{0}=1$ for $0<\beta <1$ and $n_{0}=0$ for $-1<\beta <0$. The
summation formula more adapted for application to (\ref{G2}) is obtained
from (\ref{SF2}):%
\begin{equation}
\sum_{n=-\infty }^{\infty }f(n-\beta )=\int_{-\infty }^{\infty }{dx\,f(x)+}%
i\int_{0}^{\infty }{dx\,}\sum_{s=\pm 1}s{\frac{f(e^{si\pi /2}x)-f(-e^{-si\pi
/2}x)}{e^{2\pi \left( x-si\beta \right) }-1}.}  \label{SF3}
\end{equation}

For the summation of the series in (\ref{G2}) we take%
\begin{equation}
f(z)=\frac{\cos (\sqrt{z^{2}/L^{2}+E_{D-1}^{2}}\Delta t)}{\sqrt{%
z^{2}/L^{2}+E_{D-1}^{2}}}e^{iz\Delta x^{D}/L},  \label{fG}
\end{equation}%
with $E_{D-1}=\sqrt{\mathbf{p}_{D-1}^{2}+m^{2}}$. For this function, the
contribution of the part with the integral $\int_{-\infty }^{\infty }{%
dx\,f(x)}$ coincides with the HF (\ref{G0}) and the topological contribution
is given by last term in (\ref{SF3}). For the function (\ref{fG}) one has $%
f(e^{si\pi /2}x)-f(-e^{-si\pi /2}x)=0$ for $x<LE_{D-1}$. Introducing a new
integration variable $y=\sqrt{x^{2}/L^{2}-E_{D-1}^{2}}$, for the difference $%
G_{\mathrm{s}}(x,x^{\prime })=G(x,x^{\prime })-G_{0}(x,x^{\prime })$ we get%
\begin{equation*}
G_{\mathrm{s}}(x,x^{\prime })=2\int d\mathbf{p}_{\perp }\frac{e^{i\mathbf{p}%
_{\perp }\cdot \Delta \mathbf{x}_{\perp }}}{(2\pi )^{D}}\int_{0}^{\infty }%
\frac{{dy}}{u}{\,\cosh (y\Delta t)\,}\sum_{s=\pm 1}{\frac{e^{-su\Delta x^{D}}%
}{e^{2\pi \left( Lu+si\alpha \right) }-1},}
\end{equation*}%
where $u=\sqrt{y^{2}+\mathbf{p}_{\perp }^{2}+m^{2}}$. Integrating over the
angular part of the integral over $\mathbf{p}_{\perp }$ and using the
expansion $1/(e^{w}-1)=\sum_{n=1}^{\infty }e^{-nw}$, we then introduce polar
coordinates in the plane $(y,|\mathbf{p}_{\perp }|)$. The integral over the
radial coordinate $\sqrt{y^{2}+\mathbf{p}_{\perp }^{2}}$ is evaluated by
using the formula from \cite{Prud86} and is expressed in terms of the
modified Bessel function $K_{\nu }(z)$. The final expression reads%
\begin{equation}
G_{\mathrm{s}}(x,x^{\prime })=\frac{2m^{D-1}}{(2\pi )^{\frac{D+1}{2}}}%
\sum_{n=-\infty }^{+\infty ^{\prime }}e^{2i\pi n\alpha }f_{\frac{D-1}{2}%
}\left( m\sqrt{|\Delta \mathbf{x}_{\perp }|^{2}+\left( \Delta x^{D}-2\pi
nL\right) ^{2}-\left( \Delta t\right) ^{2}}\right) ,  \label{Gsub1}
\end{equation}%
where we have defined the function $f_{\nu }(z)=K_{\nu }(z)/z^{\nu }$ and
the prime on the summation sign means that the term $n=0$ should be excluded
from summation (it can be shown that the term with $n=0$ coincides with the
function $G_{0}(x,x^{\prime })$). Hence, the application of the summation
formula (\ref{SF3}) allowed us to extract explicitly the topological
contribution.

The topological part (\ref{Gsub1}) can be used for the evaluation of the
topological contributions in the VEVs of the field squared, current density
and energy-momentum tensor. The VEV of the field squared is directly
obtained from (\ref{Gsub1}) in the coincidence limit $x^{\prime }\rightarrow
x$:%
\begin{equation}
\left\langle \varphi \varphi ^{\dagger }\right\rangle _{\mathrm{t}}=\frac{%
2m^{D-1}}{(2\pi )^{\frac{D+1}{2}}}\sum_{n=1}^{\infty }\cos \left( 2\pi
n\alpha \right) f_{\frac{D-1}{2}}\left( 2\pi nmL\right) .  \label{phi2}
\end{equation}%
Note that in the Minkowski spacetime with trivial spatial topology $R^{D}$
the VEV of the field squared is renormalized to zero and $\left\langle
\varphi \varphi ^{\dagger }\right\rangle _{\mathrm{t}}$ presents the
renormalized VEV in the topology $R^{D-1}\times S^{1}$. The vacuum current
density is obtained from the HF by using the relation $\left\langle
j_{l}\right\rangle =e\lim_{x^{\prime }\rightarrow x}(\partial _{l}-\partial
_{l}^{\prime })G_{\mathrm{s}}(x,x^{\prime })/2$, where $e$ is the charge of
the field quanta. The charge density and the components of the current
density along uncompact dimensions are zero, $\left\langle
j_{l}\right\rangle =0$ for $l=0,1,\ldots ,D-1$. The only nonzero component
is directed along the compact dimension and the corresponding expression
reads%
\begin{equation}
\left\langle j^{D}\right\rangle =\frac{4em^{D+1}L}{(2\pi )^{\frac{D-1}{2}}}%
\sum_{n=1}^{\infty }n\sin (2\pi n\alpha )f_{\frac{D+1}{2}}(2\pi nmL).
\label{jrT0}
\end{equation}%
In the geometry with decompactified direction $x^{D}$ the current density
vanishes by the symmetry and the VEV $\left\langle j^{D}\right\rangle $
coincides with the total current density. The applications of the summation
formula (\ref{SF3}) to scalar and fermionic fields in more general models
with toroidal spatial dimensions at both zero and finite temperatures are
considered in \cite{Bell10}-\cite{Bell14}. The vacuum effects of toroidal
compactifications in de Sitter and anti-de Sitter bulks are studied in \cite%
{Bell13b,Beze15}.

\section{Vacuum densities for spherical boundaries}

\label{sec:Boundary}

As examples of more complicated applications of the GAPF, in this section we
consider the problems with spherical boundaries. The initial interest to the
study of the Casimir effect for a spherical boundary has been motivated by a
semiclassical model for a charged particle where the Casimir force plays the
role of the Poincar\'{e} stress that balances the Coulomb repulsive forces
to stabilize the particle. Compared to the case of planar boundaries, the
evaluation of the Casimir energy for spherical boundaries is mathematically
more difficult problem. Different methods have been applied, including the
generalized zeta function technique (see \cite{Most97,Bord09,Casi11}). An
efficient way for the investigation of the VEVs of local observables is
based on the use of the summation formula obtained from GAPF. For a scalar
field with Dirichlet boundary conditions on a sphere in background of the
Minkowski spacetime the eigenvalues of the radial quantum number inside the
sphere are expressed in terms of the zeros of the Bessel function $J_{\nu
}(z)$ with respect to the argument. Denoting by $z=\lambda _{\nu ,k}$, $%
k=1,2,\ldots $, those zeros in the right half-plane, the following summation
formula is obtained from the GAPF:%
\begin{eqnarray}
&&\lim_{b\rightarrow +\infty }\left[ 2\sum_{k=1}^{n}\left. \frac{f(z)}{%
zJ^{\prime }{}_{\nu }^{2}(z)}\right\vert _{z=\lambda _{\nu ,k}}-\int_{0}^{b}{%
f(x)dx}\right]   \notag \\
&&\qquad =\frac{\pi }{2}\underset{z=0}{\mathrm{Res}}\left[ f(z)\frac{Y_{\nu
}(z)}{J_{\nu }(z)}\right] -\frac{1}{\pi }\int_{0}^{\infty }dx\frac{K_{\nu
}(x)}{I_{\nu }(x)}\sum_{j=\pm }e^{-j\nu \pi i}f(xe^{j\frac{\pi i}{2}}),
\label{SFBes}
\end{eqnarray}%
where $I_{\nu }(x)$ and $K_{\nu }(x)$ are the modified Bessel functions. The
derivation of this formula for more general series over the zeros of the
combination $AJ_{\nu }(z)+BzJ_{\nu }^{\prime }(z)$ and the corresponding
conditions on the function $f(z)$ can be found in \cite{Saha08,Saha87}. In
the special case $\nu =1/2$ the standard Abel-Plana formula is obtained from
(\ref{SFBes}). The applications of the formula (\ref{SFBes}) in the Casimir
effect for spherical and cylindrical boundaries were discussed in \cite%
{Saha01,Rome01}. Series over the zeros $\lambda _{\nu ,k}$ also appear in
the Casimir effect for parallel branes in anti-de Sitter spacetime \cite%
{Saha05,Bell16,Bell20}.

More general problem with a spherical boundary in background of constant
negative curvature space has been considered in \cite{Saha08b,Bell14b}. The
corresponding line element reads $ds^{2}=dt^{2}-a^{2}(dr^{2}+\sinh
^{2}rd\Omega _{D-1}^{2})$, where the constant $a$ determines the curvature
radius and $d\Omega _{D-1}^{2}$ is the line element on $S^{D-1}$ with unit
radius. We consider a spherical boundary with radius $r=r_{0}$ on which the
scalar field $\varphi (x)$ obeys the Dirichlet boundary condition, $\varphi
(x)|_{r=0}=0$. The radial part of the field mode functions are expressed in
terms of the function%
\begin{equation}
p_{iz-1/2}^{-\mu }(u)=\frac{P_{iz-1/2}^{-\mu }(u)}{(u^{2}-1)^{(D-2)/4}}%
,\;u=\cosh r,  \label{pemu}
\end{equation}%
where $\mu =l+D/2-1$, $l=0,1,2\ldots $, and $z$ is related to the energy $E$
of the mode by the formula $E(z)=a^{-1}\sqrt{z^{2}+z_{m}^{2}}$ with $z_{m}=%
\sqrt{m^{2}a^{2}-D(D-1)\left( \xi -\xi _{D}\right) }$. Here, $P_{\nu }^{-\mu
}(u)$ is the associated Legendre function of the first kind \cite{Abra72}
and $\xi _{D}=(D-1)/4D$ is the curvature coupling parameter for a
conformally coupled field. The eigenvalues of the quantum number $z$ inside
the sphere, $r<r_{0}$, are determined by the boundary condition and they are
roots of the equation $P_{iz-1/2}^{-\mu }(u_{0})=0$, $u_{0}=\cosh r_{0}$.
Let $z=z_{k}$, $k=1,2,\ldots $, be positive roots arranged in ascending
order of magnitude. The mode-sum for the positive frequency Wightman
function (WF) $W(x,x^{\prime })$ inside the sphere contains the series%
\begin{equation*}
\sum_{k=1}^{\infty }\left. zT_{\mu }(z)|\Gamma (\mu
+iz+1/2)|^{2}p_{iz-1/2}^{-\mu }(u)p_{iz-1/2}^{-\mu }(u^{\prime })\frac{%
e^{-iE(z)\Delta t}}{E(z)}\right\vert _{z=z_{k}},
\end{equation*}%
where $\Gamma (x)$ is the gamma function and 
\begin{equation}
T_{\mu }(z)=\frac{Q_{iz-1/2}^{-\mu }(u_{0})}{\partial _{z}P_{iz-1/2}^{-\mu
}(u_{0})}\cos [\pi (\mu -iz)],  \label{T}
\end{equation}%
with $Q_{\nu }^{-\mu }(u)$ being the associated Legendre function of the
second kind \cite{Abra72}.

A summation formula for the series over the zeros $z_{k}$ is obtained from
the GAPF taking $f(z)=\sinh (\pi z)h(z)$ and 
\begin{equation}
g(z)=\frac{e^{i\mu \pi }h(z)}{\pi iP_{iz-1/2}^{-\mu }(u)}\sum_{j=\pm }\cos
[\pi (\mu -jiz)]Q_{jiz-1/2}^{-\mu }(u),
\end{equation}%
with a function $h(z)$ obeying the condition $|h(z)|<\varepsilon (x)e^{cy\,%
\mathrm{arccosh\,}u}$ for $|z|\rightarrow \infty $, where $c<2$ and $%
\varepsilon (x)e^{\pi x}\rightarrow 0$ for $x\rightarrow +\infty $. The
summation formula reads \cite{Saha08b}%
\begin{eqnarray}
&&\sum_{k=1}^{\infty }T_{\mu }(z_{k},u)h(z_{k})=\frac{e^{-i\mu \pi }}{2}%
\int_{0}^{\infty }dx\,\sinh (\pi x)h(x)  \notag \\
&&\qquad -\frac{1}{2\pi }\int_{0}^{\infty }dx\,\frac{Q_{x-1/2}^{-\mu }(u)}{%
P_{x-1/2}^{-\mu }(u)}\cos [\pi (\mu +x)]\sum_{j=\pm }h(xe^{j\pi i/2}).
\label{SFleg}
\end{eqnarray}%
The contribution of the first integral in the right-hand side of (\ref{SFleg}%
) to the WF coincides with the corresponding function $W_{0}(x,x^{\prime })$
in the boundary-free geometry. As a result, the WF is decomposed as $%
W(x,x^{\prime })=W_{0}(x,x^{\prime })+W_{\mathrm{s}}(x,x^{\prime })$, where
the sphere induced contribution is obtained from the second integral in (\ref%
{SFleg}). Though the coincidence limit $x^{\prime }\rightarrow x$ is
divergent for separate parts $W(x,x^{\prime })$ and $W_{0}(x,x^{\prime })$,
the difference is finite for $r<r_{0}$. The boundary induced part in the VEV
of the field squared is directly obtained from $W_{\mathrm{s}}(x,x^{\prime })
$: $\left\langle \varphi ^{2}\right\rangle _{\mathrm{s}}=\lim_{x^{\prime
}\rightarrow x}W_{\mathrm{s}}(x,x^{\prime })$. The corresponding expression
reads%
\begin{equation}
\langle \varphi ^{2}\rangle _{\mathrm{s}}=-\frac{\Gamma (D/2)}{\pi ^{\frac{D%
}{2}+1}a^{D-1}}\sum_{l=0}^{\infty }\frac{\mu \Gamma (l+D-2)}{\Gamma
(D-1)l!e^{i\mu \pi }}\int_{z_{m}}^{\infty }dz\,z\frac{Q_{z-1/2}^{\mu }(u_{0})%
}{P_{z-1/2}^{-\mu }(u_{0})}\frac{[p_{z-1/2}^{-\mu }(u)]^{2}}{\sqrt{%
z^{2}-z_{m}^{2}}}.  \label{phi2s}
\end{equation}%
The sphere induced contribution in the VEV of the energy-momentum tensor is
obtained acting on the function $W_{\mathrm{s}}(x,x^{\prime })$ by the
corresponding differential operator $\hat{T}_{ik}(x,x^{\prime })$ and taking
the limit $x^{\prime }\rightarrow x$.

Series over the zeros $z_{k}$ appear also in the evaluation of the VEVs
inside a sphere in de Sitter spacetime foliated by negative curvature
spatial sections. The corresponding metric tensor is given by the line
element 
\begin{equation}
ds^{2}=dt^{2}-b^{2}\sinh ^{2}\left( t/b\right) (dr^{2}+\sinh ^{2}rd\Omega
_{D-1}^{2}),\;t\in \lbrack 0,\infty ),  \label{ds2ds}
\end{equation}%
where the constant $b$ determines the spacetime curvature radius and the
Ricci scalar is expressed as $R=D\left( D+1\right) /b^{2}$. Again, the
radial part of the mode function is given by the function (\ref{pemu}) and
for Dirichlet boundary condition on the sphere of radius $r_{0}$ the
eigenmodes of $z$ are zeros of the function $P_{iz-1/2}^{-\mu }(u_{0})$. The
series in the corresponding mode sum for the WF has the form%
\begin{eqnarray}
&&\sum_{k=1}^{\infty }zT_{\mu }\left( z\right) P_{\nu -1/2}^{iz}\left( \cosh
(t/b)\right) P_{\nu -1/2}^{-iz}\left( \cosh (t^{\prime }/b)\right)   \notag
\\
&&\;\times \left. \frac{\left\vert \Gamma \left( \mu +iz+1/2\right)
\right\vert ^{2}}{\sinh \left( \pi z\right) }p_{iz-1/2}^{-\mu }\left(
u\right) p_{iz-1/2}^{-\mu }\left( u^{\prime }\right) \right\vert _{z=z_{k}},
\label{Ser2}
\end{eqnarray}%
where $\nu =\sqrt{D^{2}/4-\xi D\left( D+1\right) -m^{2}b^{2}}$. The
application of the summation formula (\ref{SFleg}) to this series allows to
separate the sphere induced contribution and to find the corresponding VEVs
directly in the coincidence limit. The investigation of the VEVs in both the
interior and exterior region for a more general Robin boundary condition is
presented in \cite{Saha21}.

\section{Summary}

\label{sec:Sum}

We have considered some applications of the GAPF in the investigations of
the topology and boundary induced effects in quantum field theory. Summation
formule are obtained from the GAPF which allow to separate explicitly the
effects induced by topology and boundaries in the expectation values of
local physical observables and to present those contributions in terms of
strongly convergent integrals for points away from boundaries. Another
advantage is that in the corresponding representations the explicit
knowledge of the eigenmodes of quantum numbers is not required. Other
applications can be found in \cite{Saha08}.

\section*{Acknowledgements}

The work was supported by the grant No. 21AG-1C047 of the Higher Education
and Science Committee of the Ministry of Education, Science, Culture and
Sport RA.

\end{document}